\documentclass[9pt,conference]{IEEEtran}

\usepackage[latin1]{inputenc}
\usepackage[cmex10]{amsmath}
\usepackage{color}
\usepackage{graphicx}
\usepackage{amssymb}
\usepackage{cite}
\usepackage{amssymb}

\begin{document}

\title{Sliding-Trellis based Frame Synchronization}

\author{Usman ALI$^{1}$, Michel KIEFFER$^{1,2}$, and Pierre DUHAMEL$^{1}$\\
 {$^{1}$ L2S, CNRS -- SUPELEC -- Univ Paris-Sud - 91192 Gif-sur-Yvette,
France}\\
 $^{2}$ on sabbatical leave at LTCI, CNRS -- Télécom ParisTech -
75013 Paris, France}

\maketitle
\begin{abstract}
Frame Synchronization (FS) is required in several communication standards
in order to recover the individual frames that have been aggregated
in a \textit{burst}. This paper proposes a low-delay and reduced-complexity
\textit{Sliding Trellis} (ST)-based FS technique, compared to our
previously proposed trellis-based FS method. Each burst is divided
into overlapping windows in which FS is performed. Useful information
is propagated from one window to the next. The proposed method makes
use of soft information provided by the channel, but also of all sources
of redundancy present in the protocol stack. An illustration of our
ST-based approach for the WiMAX Media Access Control (MAC) layer is
provided. When FS is performed on bursts transmitted over Rayleigh
fading channel, the ST-based approach reduces the FS latency and complexity
at the cost of a very small performance degradation compared to our
full complexity trellis-based FS and outperforms state-of-the-art
FS techniques. 
\end{abstract}
\begin{IEEEkeywords} CRC, Cross-layer decoding, Frame synchronization,
Joint decoding, Segmentation. \end{IEEEkeywords}

\section{Introduction}

FS is an important problem arising at various layers of the protocol
stack of several communication systems. The most obvious being, at
Physical (PHY) layer, to recover the payload and the side information
(headers, etc...) of PHY packets or \emph{frames}. First results \cite{Barker1953,Massey72}
considered data streams in which regularly spaced fixed patterns or
\emph{Synchronization Words} (SW) are inserted to delimit fixed-length
frames, as is the case, \emph{e.g.}, at the PHY layer of Digital Video
Broadcasting - Handheld (DVB-H) for MPEG2 transport stream frames.
Synchronization tools based on the maximization of the correlation
between the SW and the received data have been proposed in \cite{Barker1953}.
This has been improved in \cite{Massey72}, where the optimal statistic
for FS has been proposed for the AWGN case, taking into account the
presence of data around the SW. This was further extended in \cite{Lui1987}
for more sophisticated transmission schemes and in \cite{Choi2002,Lee2009}
to provide robustness against frequency and phase errors.

In many communication systems, frames are of variable-length, see,
\emph{e.g.}, the 802.11/802.16 standards \cite{IEEE802.11n,IEEE_802.16_2004}
for Wireless Local Area Networks (WLANs). In absence of SW, the Header
Error Control (HEC) field of the header has been employed in \cite{Ueda2001}
to perform FS with an automaton adapted from \cite{ITU.432-1}. A
length field, assumed present in the frame header, facilitates the
FS when the noise is moderate. In \cite{Chiani2006,Maria2009}, several
hypothesis testing techniques have been proposed to perform FS in
presence of SW, which are further extended in \cite{Maria2010} to
exploit \textit{a priori} information on the prevalence of ones and
zeros in the payload at the price of a small additional signaling
and computational complexity. However, insertion of SW requires some
modification to the transmitter, which might not be standard compliant.
Most of these FS techniques work \emph{on-the-fly}, \emph{i.e.}, at
each time, only few data samples are processed to perform FS and almost
no latency is introduced.

Initially, FS has mainly been considered at PHY layer, albeit this
problem may also occur at upper layers of the protocol stack. In fact,
some frame aggregation techniques at intermediate protocol layers
have been proposed recently in order to reduce the signalization overhead,
see, \emph{e.g.}, \cite{Sidelnikov2006} in the context of 802.11
standard. Efficient FS in this context is thus very important, since
if some frames are not correctly delineated, a large amount of bits
has to be retransmitted. In this context, processing %using a \emph{hold-and-sync(hronize)} approach, where 
a whole \textit{\emph{burst}} %is processed 
at each step may significantly improve the FS performance. This was
evidenced in \cite{Ali2009} for the segmentation of MAC frames aggregated
in WiMAX PHY \emph{burst}s, where \emph{Joint Protocol-Channel
Decoding (JPCD)} is performed using a modified BCJR algorithm \cite{Bahl74}
to obtain the frame boundaries. It exploits all available information:
soft information at the output of the channel (or channel decoder)
as well as the structure of the protocol layers (SW, known fields
in headers, presence of Cyclic Redundancy Check (CRC) or checksums,
\emph{etc}).

This paper proposes an adaptation of the reduced-complexity \textit{Sliding
Window} (SW) \cite{Benedetto96} variant of the BCJR algorithm, presented
for the decoding of convolutional codes, to develop a low-delay and
reduced-complexity version of the FS technique presented in \cite{Ali2009}.

The FS problem is first stated as a maximum \emph{a posteriori} (MAP)
estimation problem in Section~\ref{sec:MAP_Estimation_for_FS}. Then,
Section~\ref{sec:Trellis_based_FS} reformulates the trellis-based
technique for FS introduced in \cite{Ali2009}. The proposed \emph{ST}-based
algorithm is presented in Section~\ref{sec:Overlapped_BCJR} and
is illustrated in Section~\ref{sec:Simulationresults} with the FS
of WiMAX MAC frames aggregated in \textit{\emph{bursts}}.

\section{MAP estimation for Frame Synchronization\label{sec:MAP_Estimation_for_FS}}

\subsection{Frame structure \label{sub:Structure_packet}}

Consider the $n$-th variable-length frame at a given protocol layer.
This frame is assumed to contain $\lambda_{n}=\ell_{h}+\ell_{p,n}$
bits, where the leading $\ell_{h}$ bits represent the frame header,
of fixed length, and the remaining $\ell_{p,n}$ bits constitute the
variable-length payload. In the header, $\ell_{c}$ bits are some
HEC bits: CRC or checksum. The length $\lambda_{n}$ is assumed to
be a realization of a stationary memoryless process $\Lambda$ characterized
by \begin{equation}
\pi_{\lambda}=\Pr\left(\Lambda=\lambda\right)\neq0\text{ for }\ell_{\min}\leqslant\lambda\leqslant\ell_{\max},\label{Eq:LengthConstraint}\end{equation}
 where $\ell_{\min}$ and $\ell_{\max}$ are the minimum and maximum
length in bits of a frame.

The header $\mathbf{h}_{n}$ of the $n$-th frame can be partitioned
into four fields. The \emph{constant} field $\mathbf{k}$, contains
all bits which do not change from one frame to the next. It includes
the SW indicating the beginning of the frame, and other bits which
remain constant \cite{Marin2008} once the communication is established.
The header is assumed to contain a \emph{length} field $\mathbf{u}_{n}$,
indicating the size of the frame in bits $\lambda_{n}$, including
the header. Our task is to estimate the successive values taken by
this quantity in all frames of the burst. The \emph{other} field $\mathbf{o}_{n}$,
gathers all bits of the header which are not %fully determined and will not be 
used to perform FS. Finally, the \textit{\emph{HEC}} field $\mathbf{c}_{n}$
is assumed to cover the $\ell_{h}-\ell_{c}$ \char`\"{}working\char`\"{}
bits of the header, \emph{i.e.}, $\mathbf{c}_{n}=\mathbf{f\left(\mathbf{k},\mathbf{u}_{n},\mathbf{o}\right)}$,
where $\mathbf{f}$ is some (CRC or checksum) encoding function. The
payload (assumed not protected by the HEC field) of the $n$-th frame
is denoted by $\mathbf{p}_{n}$. It is modeled here as a binary symmetric
sequence.

In what follows, the length of a vector $\mathbf{z}$ (in bits) is
denoted as $\ell\left(\mathbf{z}\right)$ and its observation (soft
information) provided either by a channel, a channel decoder, or a
lower protocol layer is denoted as $\mathbf{y}_{z}$. $\mathbf{z}_{a}^{b}$
represents the sub-vector of $\mathbf{z}$ between indexes $a$ and
$b$ (in bits).

\subsection{Aggregated frames within a burst}

Consider a burst of $L$ bits consisting of $N$ \emph{aggregated
frames}. This burst contains either $N-1$ \emph{data} frames and
an additional \emph{padding} frame containing only padding bits, or
$N$ data frames. Assume that each of these frames, except the padding
frame, contains a header and a payload and follows the same syntax,
as described in Section~\ref{sub:Structure_packet}.

\begin{figure}[btp]
 \centering 
 \includegraphics[width=0.8\columnwidth]{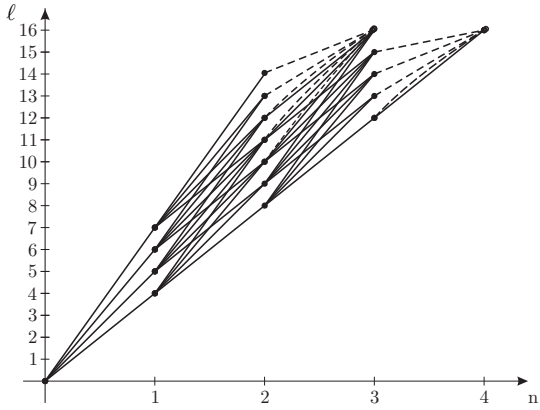}

 \caption{$L=16$ bits with $\ell_{\min}=4$ bits and $\ell_{\max}=7$ bits}
\label{fig:WimaxPHYTrellis} 
\end{figure}

Assuming that $L$ is fixed \emph{before} frame aggregation and that
$N$ is not determined \emph{a priori}, the accumulated length in
bits $\ell$ of the $n$ first aggregated frames can be described
by a Markov process, which state is denoted by $S_{n}$. With this
representation, the successive values taken by $S_{n}$, for $n=0,1,\dots$
can be described by a trellis \cite{Ali2009} such as that of Figure~\ref{fig:WimaxPHYTrellis},
with \emph{a priori} state transition probabilities $p\left(S_{n}=\ell\,|\, S_{n-1}=\ell^{\prime}\right)$
deduced from \eqref{Eq:LengthConstraint}. If $\ell<L$, then 
\begin{equation}
P\left(S_{n}=\ell\,|\, S_{n-1}=\ell^{\prime}\right)=\left\{ \begin{array}{ll}
\pi_{\ell-\ell^{\prime}} & \text{if }\ell_{\min}\leqslant\ell-\ell^{\prime}\leqslant\ell_{\max}\\
0 & \text{else,}\end{array}\right.\label{eq:MarkovMod1}
\end{equation}
and if $\ell=L$, then 
\begin{eqnarray}
P\left(S_{n}=L\,|\, S_{n-1}=\ell^{\prime}\right) & = & \left\{ \begin{array}{l}
0\text{, if }L-\ell^{\prime}>\ell_{\max}\\
1\text{, if }0<L-\ell^{\prime}<\ell_{\min}\\
\sum\limits _{k=L-\ell'}^{\ell_{\max}}\pi_{k}\text{, else.}\end{array}\right.
\label{eq:MarkovMod2}
\end{eqnarray}
In the trellis, dashed transitions correspond to padding frames and
plain transitions correspond to data frames.

\subsection{Estimators for the number of frames and their boundaries}

Consider a \textit{burst} $\mathbf{x}_{1}^{L}$ of $N$ aggregated
frames and some vector $\mathbf{y}_{1}^{L}$ containing soft information
about the bits of $\mathbf{x}_{1}^{L}$. Here, one assumes that the
first entry of $\mathbf{y}_{1}^{L}$ corresponds to the first bit
of $\mathbf{x}_{1}^{L}$. This assumption is further discussed in
Section~\ref{sub:Limitations}.

The suboptimal MAP estimator presented in \cite{Ali2009} consists
in first estimating $N$ and then in estimating the locations of the
beginning and of the end of the frames. The MAP estimate $\widehat{N}_{\text{MAP}}$
of $N$ is given by 
\begin{equation}
\widehat{N}_{\text{MAP}}=\arg\max_{N_{\min}\leqslant n\leqslant N_{\max}}P\left(S_{n}=L|\mathbf{y}_{1}^{L}\right),\label{Eq:MAPN}\end{equation}
 with $N_{\min}=\left\lceil L/\ell_{\max}\right\rceil $, $N_{\max}=\left\lceil L/\ell_{\min}\right\rceil $,
and $\left\lceil \cdot\right\rceil $ denoting the upward rounding.
Once $\widehat{N}_{\text{MAP}}$ is obtained, the MAP estimate for
the index $\ell_{n}$ of the last bit of the $n$-th $(n=1,...,\widehat{N}_{\text{MAP}})$
frame is 
\begin{equation}
\widehat{\ell}_{n}=\arg\max_{\ell}P\left(S_{n}=\ell|\mathbf{y}_{1}^{L}\right),\label{Eq:MAPEpsilon}\end{equation}
 and the length $\lambda_{n}$ of the $n$-th frame is estimated as\begin{equation}
\widehat{\lambda}_{n}=\arg\max_{\ell}P\left(S_{n}=\ell|\mathbf{y}_{1}^{L}\right)-\arg\max_{\ell}P\left(S_{n-1}=\ell|\mathbf{y}_{1}^{L}\right).
\label{eq:MAPLambda}
\end{equation}

\section{Trellis-based FS Algorithm}

\label{sec:Trellis_based_FS}

In $\eqref{Eq:MAPN},$ \eqref{Eq:MAPEpsilon}, and $\eqref{eq:MAPLambda}$,
one has to evaluate $P(S_{n}=\ell|\mathbf{y}_{1}^{L})$ for all possible
values of $n$ and $\ell$. This can be performed efficiently using
the BCJR algorithm \cite{Bahl74}, by evaluating first \begin{equation}
P(S_{n}=\ell,\mathbf{y}_{1}^{L})=\alpha_{n}\left(\ell\right)\beta_{n}\left(\ell\right)\label{Eq:BCJRDec}\end{equation}
 where\begin{eqnarray}
\alpha_{n}(\ell)=P(S_{n}=\ell,\mathbf{y}_{1}^{\ell})=\sum_{\ell^{\prime}}\alpha_{n-1}(\ell^{\prime})\gamma_{n}(\ell^{\prime},\ell),\\
\beta_{n}(\ell)=P(\mathbf{y}_{\ell+1}^{L}|S_{n}=\ell)=\sum_{\ell^{\prime}}\beta_{n+1}(\ell^{\prime})\gamma_{n+1}(\ell,\ell^{\prime}),\end{eqnarray}
 and \begin{equation}
\gamma_{n}\left(\ell^{\prime},\ell\right)=P(S_{n}=\ell,\mathbf{y}_{\ell^{\prime}+1}^{\ell}|S_{n-1}=\ell^{\prime}).\end{equation}

Classical BCJR forward and backward recursions allow to evaluate $\alpha$
and $\beta$. The initial value $S_{0}$ of the forward recursion
is known, leading to $\alpha_{0}(\ell=0)=1$ and $\alpha_{0}(\ell\neq0)=0$.
For the backward recursion, assuming that all %$N_{\max}-N_{\min}+1$ 
allowed final states are equally likely (which is a quite coarse approximation),
one gets \begin{equation}
\beta_{n}(L)=\frac{1}{N_{\max}-N_{\min}+1}\text{, }N_{\min}\leqslant n\leqslant N_{\max}\text{.}\label{eq:InitBeta}\end{equation}
 All other values of $\beta_{n}(\ell)$, for $\ell<L$ are initialized
to $0$.

\subsection{Evaluation of $\gamma_{n}$\label{sub:Evaluation_gamma}}

Two cases have to be considered for evaluating $\gamma_{n}\left(\ell^{\prime},\ell\right)$.

When $\ell<L$, the transition corresponding to the $n$-th frame
cannot be the last one, thus corresponds to a data frame. Assuming
that $\ell_{\min}\leqslant\ell-\ell^{\prime}\leqslant\ell_{\max}$,
the bits between $\ell^{\prime}+1$ and $\ell$ may be interpreted
as $\mathbf{x}_{\ell^{\prime}+1}^{\ell}=\left[\mathbf{\mathbf{k}},\mathbf{\mathbf{u}}_{n},\mathbf{\mathbf{o}},\mathbf{\mathbf{c}},\mathbf{p}\right]$,
where $\mathbf{\mathbf{u}}_{n}=\mathbf{u}\left(\ell-\ell^{\prime}\right)$
is the binary representation of $\ell-\ell^{\prime}$. The corresponding
observation can be written as $\mathbf{y}_{\ell^{\prime}+1}^{\ell}=\left[\mathbf{y}_{k},\mathbf{y}_{u},\mathbf{y}_{o},\mathbf{y}_{c},\mathbf{y}_{p}\right]$.
With these notations, for $\ell\neq L$, $\gamma_{n}\left(\ell^{\prime},\ell\right)=\gamma_{n}^{\text{d}}\left(\ell^{\prime},\ell\right)$,
with\begin{eqnarray}
\gamma_{n}^{\text{d}}\left(\ell^{\prime},\ell\right) & = & p\left(S_{n}=\ell|S_{n-1}=\ell^{\prime}\right)\varphi^{\mathrm{d}}\left(\mathbf{y}_{\ell^{\prime}+1}^{\ell},\mathbf{x}_{\ell^{\prime}+1}^{\ell}\right),\label{Eq:Gamma_d}\end{eqnarray}
 where \begin{align}
\varphi^{\mathrm{d}}\left(\mathbf{y}_{\ell^{\prime}+1}^{\ell},\mathbf{x}_{\ell^{\prime}+1}^{\ell}\right) & =P\left(\mathbf{y}_{k}|\mathbf{k}\right)P\left(\mathbf{y}_{u}|\mathbf{u}\left(\ell-\ell^{\prime}\right)\right)\nonumber \\
 & \hspace{-0.5cm}\sum_{\mathbf{o}}P\left(\mathbf{y}_{o}|\mathbf{o}\right)P\left(\mathbf{y}_{c}|\mathbf{c=f}\left(\mathbf{k},\mathbf{u}\left(\ell-\ell^{\prime}\right),\mathbf{o}\right)\right)P\left(\mathbf{o}\right)\nonumber \\
 & \hspace{-0.5cm}\sum_{\mathbf{p}}P\left(\mathbf{y}_{p}|\mathbf{p}\right)P\left(\mathbf{p}\right).\label{Eq:BigSumSimplified}\end{align}
 Under the assumptions described above and for a memoryless AWGN channel with
variance $\sigma^{2}$, we get \[
P\left(\mathbf{y}_{u}|\mathbf{u}\left(\ell-\ell^{\prime}\right)\right)=P\left(\mathbf{y}_{u}|\mathbf{u}_{n}\right)=\prod_{i=1}^{\ell\left(\mathbf{u}_{n}\right)}\frac{1}{\sqrt{2\pi}\sigma}e^{-\left(\mathbf{y}_{u}(i)-\mathbf{u}_{n}(i)\right)^{2}/2\sigma^{2}}.\]
 Assuming that the values taken by $\mathbf{p}$ are all equally likely,
%$ P\left(\mathbf{y}_{u}|\mathbf{u}\left(\ell-\ell^{\prime}\right)
one gets $P\left(\mathbf{p}\right)=2^{-\ell\left(\mathbf{p}\right)}$.

In $\left(\ref{Eq:BigSumSimplified}\right)$, the sum over all possible
$\mathbf{o}$ can be evaluated with a complexity $\mathcal{O}(\ell\left(\mathbf{o}\right)2^{\ell\left(\mathbf{c}\right)})$,
as proposed in \cite{Marin2008}, see also \cite{Ali2009} for more
details.

When $\ell=L$ and $L-\ell'<\ell_{\max}$, the $n$-th frame is the
last one, and $\mathbf{x}_{\ell^{\prime}+1}^{L}=\mathbf{1}$ has also
to be considered in $\gamma_{n}\left(\ell^{\prime},L\right)$, leading
to \begin{align}
\gamma_{n}\left(\ell^{\prime},L\right) & =\sum_{p=0,1}P(S_{n}=L,\mathbf{y}_{\ell^{\prime}+1}^{L},P_{n}=p|S_{n-1}=\ell^{\prime})\nonumber \\
 & =\gamma_{n}^{\text{d}}\left(\ell^{\prime},L\right)P\left(P_{n}=0|S_{n-1}=\ell^{\prime}\right)\nonumber \\
 & +\gamma_{n}^{\text{p}}\left(\ell^{\prime},L\right)P\left(P_{n}=1|S_{n-1}=\ell^{\prime}\right),\label{eq:GammaLast}\end{align}
 where $P_{n}$ is a random variable indicating whether the $n$-th
frame is a padding frame and its \emph{a priori} probability is given
by

\begin{eqnarray}
P\left(P_{n}=1|S_{n-1}=\ell^{\prime}\right) & = & \left\{ \begin{array}{l}
0\text{, if }L-\ell^{\prime}\geq\ell_{\max}\\
1\text{, if }0<L-\ell^{\prime}<\ell_{\min}\\
\sum\limits _{\lambda=L-\ell^{\prime}+1}^{\ell_{\max}}\pi_{\lambda}\text{, else.}\end{array}\right.\label{eq:eq:Conditional_apriori_padding}\end{eqnarray}

In \eqref{eq:GammaLast}, \begin{eqnarray}
\gamma_{n}^{\text{p}}\left(\ell^{\prime},\ell\right) & = & P\left(S_{n}=L,\mathbf{y}_{\ell^{\prime}+1}^{L}|S_{n-1}=\ell^{\prime},P_{n}=1\right)\nonumber \\
 & = & P\left(\mathbf{y}_{\ell^{\prime}+1}^{L}|P_{n}=1,S_{n-1}=\ell^{\prime},S_{n}=\ell\right)\label{Eq:GammaPadding}\end{eqnarray}
 accounts for the padding frame, while
\begin{eqnarray*}
\gamma_{n}^{\text{d}}\left(\ell^{\prime},L\right) & = & P\left(S_{n}=L,\mathbf{y}_{\ell^{\prime}+1}^{\ell}|S_{n-1}=\ell^{\prime},P_{n}=0\right)\\
 & = & P\left(S_{n}=L|S_{n-1}=\ell^{\prime},P_{n}=0\right)\varphi^{\mathrm{d}}\left(\mathbf{y}_{\ell^{\prime}+1}^{\ell},\mathbf{x}_{\ell^{\prime}+1}^{\ell}\right)
\end{eqnarray*}
 accounts for the data frame, with \[
P\left(S_{n}=L|S_{n-1}=\ell^{\prime},P_{n}=0\right)=\frac{\pi_{L-\ell^{\prime}}}{\sum\limits _{\lambda=\ell_{\min}}^{L-\ell^{\prime}}\pi_{\lambda}}.\]

\subsection{Complexity evaluation\label{sub:Complexity-evaluation}}

The complexity of the FS algorithm described in Section~\ref{sec:MAP_Estimation_for_FS}
is proportional to the number of nodes or to the number of transitions
within the trellis on which FS is performed. From Figure~\ref{fig:WimaxPHYTrellis},
one sees that the trellis is lower-bounded by the line $\ell=n\ell_{\min}$
and upper-bounded by the lines $\ell=n\ell_{\max}$ and $\ell=L$.
This region may be divided into two triangular sub-regions, one with
$0\leq n\leq N_{\min}-1$, bounded between $\ell=n\ell_{\min}$ and
$\ell=n\ell_{\max}$, and the other with $N_{\min}\leq n\leq N_{\max}-1$,
bounded between $\ell=n\ell_{\min}$ and $\ell=L$. Thus, summing
up the number of nodes in each sub-region, one gets the number of
nodes in the trellis \begin{eqnarray}
\mathcal{N}_{{\normalcolor n}} & = & \sum_{n=0}^{N_{\min}-1}n\left(\ell_{\max}-\ell_{\min}\right)+\sum_{n=N_{\min}}^{N_{\max}}\left(L-n\ell_{\min}\right).\label{eq:NbNodes1}\end{eqnarray}
 Taking $N_{\min}\approx L/\ell_{\max}$ and $N_{\max}\approx L/\ell_{\min},$
\eqref{eq:NbNodes1} simplifies to \begin{equation}
\mathcal{\mathcal{N}}_{\mathrm{n}}=\frac{L^{2}}{2}\left(\frac{\ell_{\max}-\ell_{\min}}{\ell_{\max}\ell_{\min}}\right)=\mathcal{O}\left(L^{2}\right).\label{eq:Nodes}\end{equation}
 From each node, at most $\ell_{\max}-\ell_{\min}$ transitions may
emerge. Thus, from $\left(\ref{eq:Nodes}\right)$, the number of transitions
$\mathcal{\mathcal{N}}_{\mathrm{t}}$ may also approximated as $\mathcal{\mathcal{N}}_{\mathrm{t}}=\mathcal{O}\left(L^{2}\right).$

\subsection{Limitations}

\label{sub:Limitations}

The \textit{hold-and-sync} technique presented in \cite{Ali2009}
for performing FS is based on the knowledge of the beginning and length
of the \textit{\emph{burst}}. This requires an error-free decoding
of the headers of lower protocol layers, which contain this information.
This may be done using methods presented in \cite{Marin2008}, which
enable the lower layer to forward the \textit{\emph{burst}} to the
layer where it is processed. The main drawback of this FS technique
in terms of implementation is the increase in memory requirements
for storing soft information. This is estimated in \cite{Panza05,WooMobiCom07}
to be three to four times more. However, the plain trellis-based FS
algorithm, as described above, requires buffering the whole \textit{\emph{burst}}
which induces some buffering and processing delays proportional to
$L^{2}$, see \eqref{eq:Nodes}. To alleviate these problems, a new low-delay
and less-complex variant of the previously presented FS technique is now proposed.

\section{Sliding Trellis-based FS\label{sec:Overlapped_BCJR}}

In classical SW-BCJR methods, decoding is done within a window, which
at each step is shifted bit-by-bit \cite{Benedetto96} or by several
bits \cite{Gwak1999}. From one window to the next, the results obtained
during the forward iteration are reused, contrary to those of the
backward iteration. The number of bits the window is shifted at each
iteration determines the trade-off between complexity and efficiency.

Contrary to the trellis for a convolutional code, the trellis considered
in Figure~\ref{fig:WimaxPHYTrellis} has a variable number of states
for each value of the frame index $n$. \textit{\emph{One may apply
directly the SW ideas, but due to the increase of the size of the
trellis (at least for small values of $n$), this would still need
very large trellises to be manipulated, with an increased computation
time. Here, a}} \textit{\emph{ST-based approach is introduced: a reduced-size
trellis is considered in each decoding window. As in}} \cite{Gwak1999},
some overlapping between windows is considered, in order to allow
better reuse of already computed quantities and to allow complexity-efficiency
trade-offs. %
\begin{figure}[t]
 \centering\includegraphics[scale=0.75]{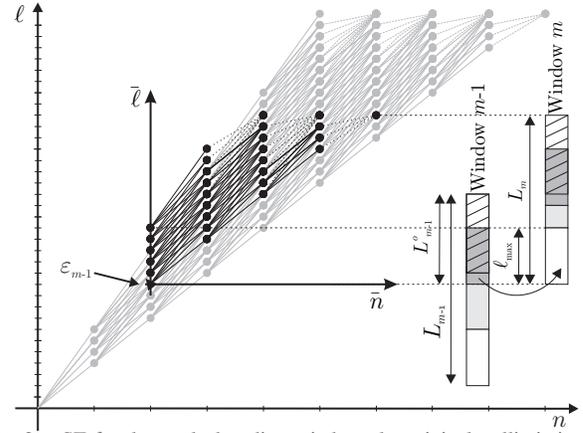} \vspace{-4mm}

\caption{ST for the $m$-th decoding window, the original trellis is in gray}

\label{fig:Sliding_Trellis1} 
\end{figure}

\subsection{Sliding Trellis\label{sub:Sliding-Trellis}}

In the proposed ST-based approach, a \textit{\emph{burst of}} $L$
bits is divided into \textcolor{black}{$M$ overlapping windows with
sizes $L_{m}$, $m=1,...,M$. For each of these windows, the bits
from} $\varepsilon_{m-1}+1$ to $\varepsilon_{m-1}+L_{m}$ are considered,
where $\varepsilon_{m-1}$ is the bit index of the last bit of the
last frame deemed reliably synchronized in the $m-1$-th window.

A ST moves from window to window to perform decoding. \textcolor{black}{One
such ST is illustrated in Figure}~\ref{fig:Sliding_Trellis1}. Let
$\bar{n}$ and $\bar{\ell}$ be the local trellis coordinates. Once
$P(S_{\bar{n}}^{m}=\bar{\ell}|\mathbf{y}_{\varepsilon_{m-1}+1}^{\varepsilon_{m-1}+L_{m}})$
is evaluated, one can apply the estimators \eqref{Eq:MAPN}, \eqref{Eq:MAPEpsilon},
and \eqref{eq:MAPLambda} to determine the number of frames $\widehat{N}_{m}$
in the $m$-th window (including the last truncated frame), the beginning,
and the length of each frame. For the $m$-th window ($m<M$), among
the $\widehat{N}_{m}$ decoded frames, only the first $\widehat{N}_{m}^{c}$
frames are considered as reliably synchronized, since enough data
and redundancy properties have been taken into account. Truncated
frames, especially when the HEC has been truncated, and the frame
immediately preceding such frames, are not considered reliable. Thus,
only the $\widehat{N}_{m}^{c}$ complete frames ending in the first
$L_{m}-\ell_{\max}-\ell_{h}$ bits of the window are considered as
reliable. The unreliable region towards the boundary of the window
is dashed in Figure~\ref{fig:Sliding_Trellis1}.

The initialization of $\beta$ is performed as in Section~\ref{sec:Trellis_based_FS},
since no knowledge from the previous window can be exploited. The
initialization of $\alpha$ and the evaluation of $\gamma$ towards
the window boundary may depend on the location of the window inside
a burst. Three types of window locations are considered: the \textit{first}
window at the start of a burst, the \emph{intermediate} windows
in the middle of the burst, and the \textit{last} window at the end
of the burst.

The first window ($m=1$) contains $L_{m}<L$ bits and starts at $\varepsilon_{0}=0$.
The decoding approach, including the initialization of $\alpha$ for
this first window is similar to that presented for the trellis-based
approach. An exception is the computation of $\gamma_{\bar{n}}\left(\bar{\ell}',\bar{\ell}\right)$,
where two cases have again to be considered. The first corresponds
to normal data frames, leading to $\gamma_{\bar{n}}^{\text{d}}\left(\bar{\ell}',L_{m}\right)$.
The second accounts for truncated data frames towards the boundary
of the window, leading to $\gamma_{\bar{n}}^{\text{t}}\left(\bar{\ell}',L_{m}\right)$,
which is detailed in Section~\ref{sub:Evaluation_gamma_sliding_trellis}.

The intermediate windows ($1<m<M$) \textcolor{black}{ contain the
bits from} $\varepsilon_{m-1}+1$ to $\varepsilon_{m-1}+L_{m}<L$,
see Figure~\ref{fig:Sliding_Trellis1}. The bit index $\varepsilon_{m-1}$,
of the last bit of the last frame deemed reliably synchronized (\textit{i.e.},
the $\widehat{N}_{m-1}^{c}$-th frame) in the $m-1$-th window, corresponds
in the local coordinates of the $m-1$-th ST to $\bar{\ell}=\varepsilon_{m-1}-\varepsilon_{m-2}$.
The $m$-th ST starts at the local coordinates $(\bar{n}=\widehat{N}_{m-1}^{c},\bar{\ell}=\varepsilon_{m-1}-\varepsilon_{m-2})$
of $m-1$-th ST. The computation of $\gamma_{\bar{n}}\left(\bar{\ell}',\bar{\ell}\right)$
for an intermediate window is identical to that of the first window.
For the initialization of $\alpha_{\bar{n}}^{m}\left(\bar{\ell}\right)$,
following the idea of the SW-BCJR decoder \cite{Benedetto1996,Benedetto96},
up to $\ell_{\max}$ initial values for $\alpha_{0}^{m}\left(\bar{\ell}\right)$
are propagated from the $m-1$-th window to the $m$-th window, see
Section~\ref{sub:InitializationAlpha_for_ST}. This allows a better
FS in case of erroneous FS in the $m-1$-th window.

The last window ($m=M$) has no incomplete frame at its end. 
Only the presence of a padding frame has to be taken
into consideration. \textcolor{black}{The decoding is performed as
in} the trellis-based approach (Section\,\ref{sec:Trellis_based_FS})\textcolor{black}{,
except for the initialization of} $\alpha_{0}^{M}\left(\bar{\ell}\right)$,
which is similar to that of the intermediate window case.

Note that the $m$-th and $m+1$-th windows overlap over $L_{m}^{o}$
bits, with $\ell_{h}+\ell_{\max}\leqslant L_{m}^{o}<\ell_{h}+2\ell_{\max}$.

\subsection{Evaluation of $\gamma_{\bar{n}}$\label{sub:Evaluation_gamma_sliding_trellis}}

When $m<M,$ transitions corresponding to truncated frames have to
be considered at the end of the window. When the size of the truncated
frame is larger than $\ell_{h}$, the header is entirely contained
in the truncated frame. In this case $\gamma_{\bar{n}}\left(\bar{\ell}',L_{m}\right)=\gamma_{\bar{n}}^{\mathrm{t}}\left(\bar{\ell}',L_{m}\right),$
with\begin{equation}
\gamma_{\bar{n}}^{\mathrm{t}}\left(\bar{\ell}',L_{m}\right)=p\left(S_{\bar{n}}^{m}=L_{m}|S_{\bar{n}-1}^{m}=\bar{\ell}'\right)\varphi^{\mathrm{t}}\left(\mathbf{y}_{\bar{\ell}'+1}^{L_{m}},\mathbf{x}_{\bar{\ell}'+1}^{L_{m}}\right).\label{eq:phit}\end{equation}
 In \eqref{eq:phit}, since truncated frames have to be considered,
$p\left(S_{\bar{n}}^{m}=L_{m}|S_{\bar{n}-1}^{m}=\bar{\ell}'\right)$
is given by \eqref{eq:MarkovMod2}. Moreover, the length of the frame,
\emph{i.e.}, the content of the length field $\mathbf{\mathbf{u}}_{n}$,
is now only known to be between $\max(L_{m}-\bar{\ell}',\ell_{\min})$
and $\ell_{\max}$ bits. Thus
\begin{align}
\varphi^{\mathrm{t}}\left(\mathbf{y}_{\bar{\ell}'+1}^{L_{m}},\mathbf{x}_{\bar{\ell}'+1}^{L_{m}}\right) & =P\left(\mathbf{y}_{k}|\mathbf{k}\right)\sum_{\mathbf{p}}P\left(\mathbf{y}_{p}|\mathbf{p}\right)P\left(\mathbf{p}\right)\nonumber \\
\sum_{\ell=\max(L_{m}-\bar{\ell}',\ell_{\min})}^{\ell=\ell_{\max}} & \vspace{-10mm}P\left(\mathbf{\mathbf{u}\left(\ell\right)}\right)\sum_{\mathbf{o}}\left(P\left(\mathbf{y}_{u}|\mathbf{u}\left(\ell\right)\right)P\left(\mathbf{y}_{o}|\mathbf{o}\right)\right.\nonumber \\
 & \left.P\left(\mathbf{y}_{c}|\mathbf{c=f}\left(\mathbf{k},\mathbf{u}\left(\ell\right),\mathbf{o}\right)\right)P\left(\mathbf{o}\right)\right).\label{eq:phitDef_modified}\end{align}
 When the size of the truncated frame is strictly less than $\ell_{h}$,
for the sake of simplicity, we assume that all bits of the truncated
header are equally likely. In such case $\gamma_{\bar{n}}\left(\bar{\ell}',L_{m}\right)=\gamma_{\bar{n}}^{\text{e}}\left(\bar{\ell}',L_{m}\right),$
with\begin{eqnarray}
\gamma_{\bar{n}}^{\mathrm{e}}\left(\bar{\ell}',L_{m}\right) & = & p\left(S_{\bar{n}}^{m}=L_{m}|S_{\bar{n}-1}^{m}=\bar{\ell}'\right)\nonumber \\
 &  & \sum_{\mathbf{x}_{\bar{\ell}'+1}^{L_{m}}}P\left(\mathbf{y}_{\bar{\ell}'+1}^{L_{m}}|\mathbf{x}_{\bar{\ell}'+1}^{L_{m}}\right)P\left(\mathbf{x}_{\bar{\ell}'+1}^{L_{m}}\right),\label{eq:gamma_incomplete_packet}\end{eqnarray}
 where $p\left(S_{\bar{n}}^{m}=L_{m}|S_{\bar{n}-1}^{m}=\bar{\ell}'\right)$
is still given by \eqref{eq:MarkovMod2} and $P(\mathbf{x}_{\bar{\ell}'+1}^{L_{m}})=2^{-\ell(\mathbf{x}_{\bar{\ell}'+1}^{L_{m}})},$
since all beginning of headers are assumed equally likely.

For $m=M,$ the evaluation of $\gamma_{\bar{n}}^{M}$ is as in Section\,\ref{sub:Evaluation_gamma}.

\subsection{Initialization of $\alpha$ in the sliding trellises\label{sub:InitializationAlpha_for_ST} }

In the SW-BCJR algorithm proposed in \cite{Benedetto96}, the $\alpha^{m}$s
evaluated in the $m$-th window are deduced from those evaluated in
the $m-1$-th window. Here, since the number of states $S_{n}$ evolves
with $n$, $\alpha_{\bar{n}}^{m}$ cannot be obtained that easily
from $\alpha_{\bar{n}}^{m-1}$.

In the $m-1$-th window, one has evaluated $\alpha_{\bar{n}}^{m-1}\left(\bar{\ell}\right)$,
with $0\leqslant\bar{\ell}\leqslant L_{m-1}$ and $0\leqslant\bar{n}\leqslant\left\lceil L_{m-1}/\ell_{\min}\right\rceil $.
We choose to propagate at most $\ell_{\max}$ values of $\alpha$
from $\bar{n}=\widehat{N}_{m-1}^{c}$ in the $m-1$-th window to $\bar{n}=0$
in the $m$-th window (for $\bar{\ell}=0,...,\ell_{\max}-1$) as follows\begin{equation}
\alpha_{0}^{m}(\bar{\ell})=\kappa\,\alpha_{\widehat{N}_{m-1}^{c}}^{m-1}\left(\varepsilon_{m-1}-\varepsilon_{m-2}+\bar{\ell}\right),\label{eq:alpha_propagation}\end{equation}
 where $\kappa$ is some normalization factor chosen such that the
$\alpha_{0}^{m}(\bar{\ell})$s sum to one. This allows the first frame
of the $m$-th window to start at any bit index between $\varepsilon_{m-1}+1$
and $\varepsilon_{m-1}+\ell_{\max}$.

\subsection{Complexity}

Consider $M$ windows of approximately the same size $L^{w}$, which
are overlapping on average on $L^{o}=\ell_{h}+1.5\ell_{\max}$~bits.
For sufficiently large $L$, one has to process $M\approx\frac{L}{L^{w}-L^{o}}$
overlapping windows, each one with \[
\mathcal{\mathcal{N}}_{\mathrm{n}}^{w}\approx\frac{\left(L^{w}\right)^{2}}{2}\left(\frac{\ell_{\max}-\ell_{\min}}{\ell_{\max}\ell_{\min}}\right)\]
 nodes. The total number of nodes to process is then

\begin{equation}
M\mathcal{\mathcal{N}}_{\mathrm{n}}^{w}\approx\frac{L}{L^{w}-L^{o}}\frac{\left(L^{w}\right)^{2}}{2}\left(\frac{\ell_{\max}-\ell_{\min}}{\ell_{\max}\ell_{\min}}\right).\label{eq:ComplexityRed}\end{equation}

This decoding complexity is smaller than that of Section~\ref{sub:Complexity-evaluation}.
A comparison for different values of burst size $L$ is provided in
Table \ref{tab:Complexity_Comparison_480} for window size $L^{w}=480+L^{o}$.
One can observe that the complexity gain increases with the size of
the burst.

\begin{table}[t]

\begin{centering}
\begin{tabular}{|l|c|c|c|c|}
\hline 
$L$ (bytes)  & 1800  & 8000  & 16000  & 24000 \tabularnewline
\hline
\hline 
Trellis-based (\# of Nodes)  & 24300  & 480000  & 1920000  & 4320000 \tabularnewline
\hline 
ST-based (\# of Nodes)  & 17400  & 77200  & 154450  & 231700 \tabularnewline
\hline 
Complexity Gain  & 1.4  & 6.2  & 12.5  & 18.6 \tabularnewline
\hline
\end{tabular}
\par\end{centering}

\caption{FS Complexity Comparison for $L^{w}=480+L^{o}$}

\label{tab:Complexity_Comparison_480} 
\end{table}

Choosing small values for $L^{w}$ reduces the latency as well as
the complexity at the cost of some sub-optimality in the decoding
performance. Note that $L^{w}$ cannot be chosen too small (smaller
than $\ell_{h}+2\ell_{\max}$) to ensure at least one reliable FS
in each window.

\section{Simulation results\label{sec:Simulationresults} }

In the WiMAX standard \cite{IEEE_802.16_2004}, the downlink (DL)
sub-frames are divided into \textit{\emph{bursts}}. Each \textit{\emph{burst}}
can contain multiple concatenated fixed-size or variable-size MAC
frames: it is filled with several MAC frames, until there is not enough
space left. Padding bytes (0xFF) are then added \cite{IEEE_802.16_2004}
at the end of the \textit{\emph{burst}}. Each MAC frame begins with
a fixed-length header, followed by a variable-length payload and ends
with an optional CRC. As we are considering only the DL case, where
the connection is already established, MAC frames belonging to a \textit{\emph{burst}}
contain only Convergence Sublayer (CS) data, so only the Generic MAC
header \cite{IEEE_802.16_2004} is possible inside a \textit{\emph{burst}}.

Some assumptions are made in what follows for the sake of simplicity.
CRC, ARQ, packing, fragmentation, and encryption are not used for
the MAC frames inside the burst. Some fields are already fixed in
a MAC header, and with the considered situation, fields such as Header
Type (HT), Encryption Control (EC), sub-headers and special payload
types (Type), Reserved (Rsv), CRC Indicator (CI), and Encryption Key
Sequence (EKS) remain constant. The LEN field, representing the length
in bytes of the MAC frame, and the Connection IDentifier (CID) have
variable contents. The Header Check Sequence (HCS), an 8-bit CRC,
is used to detect errors in the header and is a function of the content
of all header fields.

The considered simulator consists of a \textit{\emph{burst}} generator,
a BPSK modulator, a channel, and a receiver. Simulations are carried
over Rayleigh fading channels, where the modulated signal is subject
to zero mean and unit variance fast (bit) Rayleigh fading plus zero-mean
AWGN noise. For performance analysis, Erroneous Frame Location Rate
(EFLR) is evaluated as a function of the channel Signal-to-Noise Ratio
(\textit{\emph{SNR}}). It should be noted that in order to recover
a frame correctly both ends of the frame have to be correctly determined.

\textcolor{black}{In our simulations, $L=1800$ bytes. Since WiMAX
MAC frames are byte-aligned,} \textcolor{black}{\emph{i.e.}}\textcolor{black}{,
the} LEN field\textcolor{black}{{} of the frame is in bytes and all
MAC frames contain an integer number of bytes, $\alpha$, $\beta$,
and $\gamma$ are evaluated for $\ell$s corresponding to the beginning
of bytes. Data frames are randomly generated with a length uniformly
distributed between $\ell_{\min}=50$~bytes and $\ell_{\max}=200$~bytes.
If there is not enough space remaining in the burst, a padding frame
is inserted to fill the} \textit{\textcolor{black}{\emph{burst}}}\textcolor{black}{.}
\textcolor{black}{The} \textit{\textcolor{black}{\emph{burst}}} \textcolor{black}{is
then BPSK modulated and sent over the channel.}

\begin{figure}
\centering \includegraphics[scale=0.56]{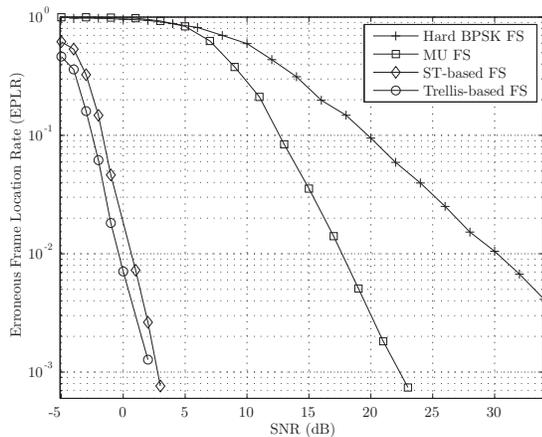}
\vspace{-4mm}

\caption{FS methods for transmission over Rayleigh channel}

\label{fig:WiMAX_MAC_Results_ray} 
\end{figure}

Simulation results for the trellis-based FS technique (Section~\ref{sec:MAP_Estimation_for_FS}),
the ST-based approach (Section~\ref{sec:Overlapped_BCJR}), a FS
based on hard decision on the received bits, and a state-of-the-art
\textcolor{black}{on-the-fly} technique (denoted by MU, for \emph{modified
Ueda's} method) described in \cite{Ueda2001}\textcolor{black}{{}
are shown in Figure~\ref{fig:WiMAX_MAC_Results_ray}.} MU technique
involves a three-state automaton for FS and uses the HCS as an error-correcting
code. Here, the method in \cite{Ueda2001} has been modified to cope
with short (8-bit) HCS, where several candidates for two-bit error
syndromes are possible.

\textcolor{black}{For the ST-based approach, the} \textit{\textcolor{black}{\emph{burst}}}\textcolor{black}{{}
is divided into three windows, with $L_{1}=600$~bytes, $L_{2}=600+L_{1}^{o}$~bytes,
and $L_{3}=600+L_{2}^{o}$~bytes. Compared to the trellis-based FS,
the ST-based approach shows a} slight performance degradation of $0.5$~dB,
but reduces the delay and the computational complexity. \textcolor{black}{On
average, the overlap is about $277$~bytes and a decrease in complexity
by a factor of 1.7 is observed}. The computational complexity of the
\textcolor{black}{on-the-fly} MU technique is much smaller than that
of the ST-based approach, at the cost of a noticeable loss in efficiency.

\section{Conclusion}

A reduced-complexity, low-delay, and efficient \textit{\emph{ST}}-based
technique for the robust FS of \textit{\emph{aggregated frames}} is
presented. Compared to the trellis-based FS algorithm proposed in
\cite{Ali2009}, the FS delay and computational complexity are significantly
reduced. The price to be paid is a slight performance degradation.

\textcolor{black}{This technique has been applied to the FS of WiMAX
MAC bursts. A significant gain in performance is obtained for Rayleigh
fading channels compared to classical FS. Extensions to perform on-the-fly}
\textcolor{black}{FS, by utilizing the features of} Ueda's method
and \textcolor{black}{at the same time exploiting the structural properties
of the headers of frames, are currently under investigation.}

\section*{Acknowledgments}

This work was partly supported by Microsoft Research through its PhD
European Scholarship program and by the NoE NEWCOM++.

%\bibliographystyle{IEEEtran}
%\bibliography{Biblio/md,Biblio/Publis_MK,Biblio/References}

% Generated by IEEEtran.bst, version: 1.12 (2007/01/11)

\end{document}